\documentclass[10pt,reprint]{iopart}
\usepackage{amsbsy}
\usepackage{amssymb}   
\usepackage{graphicx}
\usepackage{bm}
\usepackage{lineno}

\def\dd{\mbox{d}}

\def\Pt{\tilde{P}}

\def\S{{\bf S}}

\def\OO{{\mathcal O}}
\def\alphaeff{\alpha_{\mathrm{eff}}}

\begin{document}


\title[Arrival Times in Zero-Range Processes]{Arrival Times in a
Zero-Range Process with Injection and Decay}

\author{B. Hertz Shargel$^1$, M. R. D'Orsogna$^2$, T. Chou$^{1,3}$}
\address{$^{1}$Department of Mathematics, UCLA, Los Angeles, CA 90095-1555}
\address{$^{2}$Department of Mathematics, CSUN, Los Angeles, CA 91330-8313}
\address{$^{3}$Department of Biomathematics, UCLA, Los Angeles, CA 90095-1766}

\date{\today}


\begin{abstract}
Explicit expressions for arrival times of particles moving in a
one-dimensional Zero-Range Process (ZRP) are computed.  Particles are
fed into the ZRP from an injection site and can also evaporate from
anywhere in the interior of the ZRP.  Two dynamics are considered;
bulk dynamics, where particle hopping and decay is proportional to the
numqber of particles at each site, and surface dynamics, where only
the top particle at each site can hop or evaporate.  We find exact
solutions in the bulk dynamics case and for a single-site ZRP obeying
surface dynamics. For a multisite ZRP obeying surface dynamics, we
compare simulations with approximations obtained from the steady-state
limit, where mean interarrival times for both models are
equivalent. Our results highlight the competition between injection
and evaporation on the arrival times of particles to an absorbing
site.
\end{abstract}

\pacs{05.60.-k,87.16.Ac,05.10.Ln}

\vspace{5mm}

\section{Introduction}

The Zero-Range Process (ZRP) is a stochastic model to describe the
dynamics of far from equilibrium, interacting particles hopping
between lattice sites \cite{EVANS2005, Evans1, Spohn}.  The ZRP has
been used in many applications as a paradigm for transport processes,
including traffic flows, shaken granular gases, network dynamics,
phase separation, and particle condensation and clustering
\cite{Lee}. Mathematical interest also arises from the fact that a
simple connection can be made between the ZRP and the totally
asymmetric exclusion process (TASEP) \cite{EVANS2005} and that in
certain cases - particularly for conserved systems - exact
factorisable steady-state solutions can be derived \cite{Zia}.

\begin{figure}[t]
\begin{center}
\includegraphics[width=4.5in]{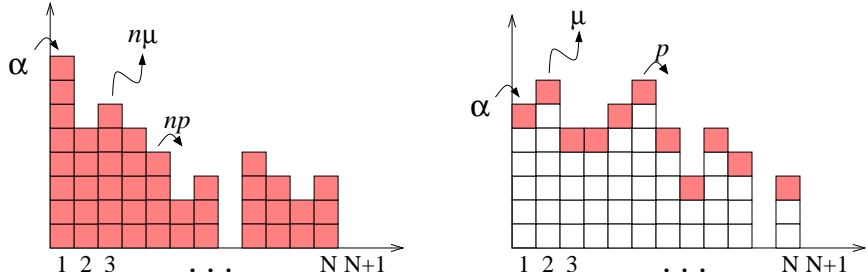}
\caption{Two realisations of a Zero-Range Process. In (a), any one of
the red particles in the bulk can hop to the right or decay, while in
(b), only the surface particle (in red) in each pile can hop or decay,
the underlying particles being protected by the top particle.}
\label{Fig1}
\end{center}
\end{figure}

In this paper, we compute the multiple passage times of particles
obeying ZRP dynamics to reach a final absorbing site. We treat a
nonconserved system where particles are injected at the origin and
evaporate as they drift right towards the end site of the lattice, as
shown in Fig. \ref{Fig1}.  This type of dynamics may be applied to
many specific micro-biological systems.  For example, molecular motors
may attach at one end of a microtubule, but desorb while traversing
it. The distribution of arrival times of the motors will depend on
their speed, and injection and desorption rates.  Other examples
include virus entry and transport to the nucleus, where the viral
cargo is transported by molecular motors while being subject to
degradation \cite{PLOSONE}, and sperm entry into egg cells, where the
first sperm to penetrate all layers of the cell triggers a block for
subsequent ones \cite{POLYSPERMY}. In all applications there is a flux
of particles, or ``immigration,'' into the first site as well as
particle annihilation at every location along the microtubule or layer
within the egg cell.

Unlike the first passage time (FPT) problem of a single, conserved
particle undergoing a simple random walk, the first passage time of a
nonconserved, multiparticle problem cannot be solved by analysing
steady-states. In problems with injection and decay, the dynamics of
particles reaching a specific ``absorbing'' site are complicated by
subsequent particle arrivals, as well as arrival times conditioned on
particles reaching the absorbing site. Since we will be concerned with
a specific initial configuration, and wish to understand how the ZRP
first reaches another configuration, we must find time-dependent
solutions for the dynamics of the ZRP.  Nonetheless, despite the
nonconserved nature of the problem, steady-state solutions can still
sometime provide useful approximations for FPTs of the ZRP in certain
limits.

We present exact solutions for arrival times in a finite-sized ZRP
obeying two specific dynamical rules.  In the first case, which we
denote as ``bulk dynamics'' and which is illustrated in Fig.\,1(a), all
particles at a site are equally likely to hop to the next site.  In
the second ``surface dynamics'' case, depicted in Figure\,1(b), only
one particle can hop to its neighboring site.  These two cases are
limits of the ZRP and may serve as a model system for many physical
systems.

We begin our analysis in the following section with bulk dynamics.  In
this case, the particles are independent and we find exact analytic
expressions for the distributions of the passage times for the $k^{\rm
th}$ particle to arrive at the absorbing site $N+1$.  For mathematical
completeness, we present two derivations of the solution. The first
involves explicit enumeration of the number of particles injected,
evaporated, and having reached the absorbing site by time $t$.  The
second involves writing down a Master equation, which is solved using
generating functions and the method of characteristics. All results
for the bulk case are exact. In the third section, we define a ZRP with
surface dynamics, where only one particle at each site, if it exists,
is allowed to hop or desorb. In this case, we can only find exact
solutions for a single site ZRP.  For a multisite ZRP, we find
particle arrival times in certain limits using a steady-state
approximation, and compare them with results derived from Monte Carlo
simulations.

\section{Zero-range Model with Bulk Dynamics}

In our bulk-dynamic ZRP, beginning at time $t=0$, particles are
injected into the first site of an empty lattice whose positions are
denoted by $\{1,2,\dots,N+1\}$.  The injection occurs as a Poisson
process with rate $\alpha$. Each particle can then hop one site to the
right with rate $p$ or evaporate with rate $\mu$, independently of
others. The forward hopping and evaporation processes continue until
the particle reaches site $N+1$ or is desorbed from the lattice.  We
wish to calculate the distribution of times for the $k^{\rm th}$
particle to reach site $N+1$.

\subsection{Explicit enumeration of particle fates} 

Denoting by $T_k$ the time at which the final absorbing site $N+1$ is
reached for the $k^{\rm th}$ time, we consider the accumulated number of
hits $H(t)$ by time $t$ defined by $P(H(t) = k) = P(T_k \leq t <
T_{k+1})$.  The primary result of this section is that $H(t)$ is
Poisson distributed, with rate parameter

\begin{eqnarray} \fl
\lambda(t) = \alpha \Bigl(\frac{p}{\mu+p}\Bigl)^N \Bigl( t -
		  \frac{N}{\mu+p} \Bigl) \nonumber \\
\label{PARAMETER} \fl
\quad \quad - \frac{\alpha}{\mu+p} \, e^{-(\mu+p)t} \, \sum_{\ell=0}^{N-1} 
\Bigl( \frac{p}{\mu+p} \Bigl)^\ell \, \sum_{m=0}^\ell \biggl( 
\frac{\mu}{p + \mu} \sum_{i=0}^m \frac{((\mu+p)t)^i}{i!}  
- \frac{((\mu+p)t)^m}{m!} \biggl) .\\
\nonumber
\end{eqnarray}
From this result we can find the survival probability $S_k(t)$
that the final site has been hit by $k-1$ or fewer particles,

\begin{eqnarray}
\label{SURVIVE}
S_k(t) = \sum_{j=0}^{k-1} P(H(t)=j) = e^{-\lambda(t)} 
\sum_{j=0}^{k-1} \frac{\lambda(t)^j}{j!}. 
\end{eqnarray}

\noindent
Similarly, the distribution of $T_i$ can be
represented as the sum

\begin{equation}
\label{T_K2}
P(T_i \leq t) = \sum_{j=i}^\infty P(H(t) = j) 
= e^{-\lambda(t)} \sum_{j=i}^\infty \frac{\lambda(t)^j}{j!}.
\end{equation}  

\noindent
To derive the distribution of $H(t)$, we begin by noting that we may
break up the event $\{H(t) = k\}$ according to how many particles $n$
were injected before time $t$, each with common probability $q(t)$ of
reaching site $N+1$ by time $t$ (see Fig. 2).  The
probability that exactly $k$ of those $n$ particles reach site $N+1$
is Binomially distributed with parameter $q(t)$, so that

\begin{eqnarray}
P(H(t) = k) 
& = \sum_{n=k}^\infty \frac{(\alpha t)^n e^{-\alpha t}}{n!} 
	\frac{n!}{k!(n-k)!} q(t)^k (1 - q(t))^{n-k} \nonumber \\
& = \frac{q(t)^k (\alpha t)^k}{k!} e^{-\alpha t}
	\sum_{n=k}^\infty \frac{(\alpha (1-q(t)) t)^{n-k}}{(n-k)!} \nonumber \\ 
\label{H_DIST}
& = \frac{(\alpha q(t) t)^k}{k!} e^{-\alpha q(t) t}.
\end{eqnarray}

\begin{figure}[t]
\begin{center}
\includegraphics[width=3.3in]{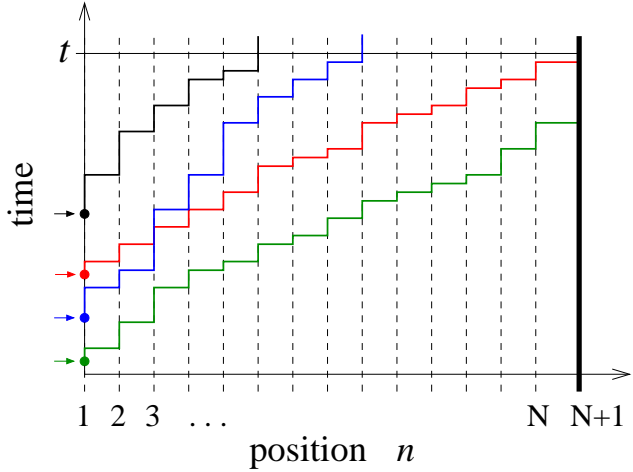}
\caption{Space-time plot of ZRP bulk dynamics.  In this realisation,
 $4$ particles are injected by time $t$, and are the only ones
capable of contributing to $H(t)$.  Trajectories that intersect the vertical
  line have arrived at site $N+1$ within time $t$. Trajectories that 
  intersect with the horizontal line at time $t$ are those that 
failed to reach the absorbing site $N+1$ by time $t$.}
\label{Fig2}
\end{center}
\end{figure}

\noindent 
This implies that $H(t)$ is Poisson distributed with parameter 

\begin{equation}
\lambda(t) = \alpha q(t) t.
\end{equation}
Deriving the probability $q(t)$ is therefore our task.  The particle
injected at time $\tau$ is characterised by its decay time
$\zeta_\tau$, which is exponentially distributed with mean $1/\mu$,
and by its arrival time $X_\tau$ at the target site $N+1$ excluding
the possibility of decay.  As this latter random variable is the sum
of exponentials, it is Gamma($N,k$)-distributed \cite{PINEDO2008}. The
probability $q(t)$ of reaching site $N+1$ is then given by the
probability that the arrival time precedes both the chosen time limit
$t$ or the time of decay, averaged over the possible injection times
$\tau$.  Symbolically,

\begin{equation}
\label{Q}
q(t) = \frac{1}{t}\int_0^t \bigl[ P(X_\tau \leq t-\tau \leq \zeta_\tau) 
	+ P(X_\tau \leq \zeta_\tau \leq t-\tau) \bigl] \dd \tau.
\end{equation}

\noindent Since $X_{\tau}$ and $\zeta_{\tau}$ are independent, the
first probability $P(X_\tau \leq t-\tau \leq \zeta_\tau) = P(X_{\tau}
< t - \tau) P(\zeta_{\tau} > t - \tau)$.  And since $\zeta_{\tau}$ is
exponentially distributed with mean $\mu$, and $X_{\tau}$ is
Gamma-distributed, the first probability in the integrand of
Eq.\,\ref{Q} is simply

\begin{eqnarray} 
P(X_\tau \leq t-\tau \leq \zeta_\tau) 
& = F_\Gamma(t-\tau;N,p)[1 - F_{\zeta_\tau}(t-\tau)] \nonumber  \\
\label{PROB1}
& = F_\Gamma(t-\tau;N,p) e^{-\mu(t-\tau)},
\end{eqnarray}

\noindent where $F_\Gamma(s;N,\beta)$ is the Gamma distribution
function.  The second probability, and many of the computations to follow,
rely on the following equivalent representations of this function,

\begin{equation}
\label{EQUIVALENCE} 
 F_\Gamma(s;N,\beta)=
\frac{\gamma(N,\beta s)}{\Gamma(N)} 
= 1 - e^{-\beta s} \sum_{\ell=0}^{N-1} \frac{(\beta s)^\ell}{\ell!}.
\end{equation}

\noindent Here $\gamma(z,w) = \int_0^w t^{z-1} e^{-t}\dd t$ is the lower
incomplete Gamma function, and the right hand equality with the Erlang
distribution holds because $N$ is an integer \cite{PINEDO2008}.  Expression
\,\ref{EQUIVALENCE} leads to the following useful identities for $F_\Gamma$:

\begin{eqnarray} \fl
\int_0^{s} F_\Gamma(u;N,\beta)\dd u 
& = \int_0^s \biggl( 1 - e^{-\beta u} \sum_{\ell=0}^{N-1} 
	\frac{(\beta u)^\ell}{\ell!} \biggl) \dd u
   = s - \sum_{\ell=0}^{N-1} \frac{1}{\ell!} \int_0^s (\beta u)^\ell e^{-\beta u}\dd u 
	\nonumber \\
\label{IDENTITY1}
& = s - \frac{1}{\beta} \sum_{\ell=0}^{N-1} 
		\frac{\gamma(\ell + 1,\beta s)}{\Gamma(\ell + 1)}
   \equiv s - \frac{1}{\beta} \sum_{\ell=0}^{N-1} F_\Gamma(s;\ell+1,\beta). \\
\nonumber
\end{eqnarray}

\noindent and

\begin{equation}
\label{IDENTITY2} \fl
\int_0^{s} e^{-\eta u} \, F_\Gamma(u;N,\beta)\dd u = \frac{1}{\eta}
\bigl(1 - e^{-\eta s} \bigl) \, - \, \frac{1}{\eta + \beta}
\sum_{\ell=0}^{N-1} \Bigl(\frac{\beta}{\eta+\beta} \Bigl)^\ell \,
F_\Gamma(s;\ell+1,\eta + \beta).
\end{equation}

\noindent Returning to the second probability in Eq.\,\ref{Q}, 
Eq.\,\ref{IDENTITY2} yields

\begin{eqnarray} \fl 
P(X_\tau \leq \zeta_\tau \leq t-\tau)
& = \int_0^{t-\tau} \mu e^{-\mu s} F_\Gamma(s;N,p)\dd s \nonumber \\
\label{SECOND}
& \hspace{-1cm} = 1 - e^{-\mu (t-\tau)} \, - \, \frac{\mu}{\mu+p} \sum_{\ell=0}^{N-1} 
	\Bigl(\frac{p}{\mu+p} \Bigl)^\ell \, F_\Gamma(t-\tau;\ell+1,\mu+p),
\end{eqnarray} 
 
\noindent 
so that upon substituting Eqs.\, \ref{PROB1}  and \ref{SECOND}
into Eq.\,\ref{Q}, we obtain

\begin{eqnarray*} \fl
q(t) = 1 + \frac{1}{t} \int_0^t \biggl[
& F_\Gamma(t-\tau,N,p) e^{-\mu(t-\tau)} - e^{-\mu(t-\tau)} \\
& - \frac{\mu}{\mu+p} \sum_{\ell=0}^{N-1} \Bigl( \frac{p}{\mu+p} \Bigl)^\ell
	F_\Gamma(t-\tau; \ell + 1, p + \mu) \biggl] d\tau.
\end{eqnarray*}

\noindent Evaluating the integral termwise using Eqs.\,\ref{IDENTITY1} and 
\ref{IDENTITY2} we find

\begin{equation*} \fl
q(t) = 1 - \frac{1}{t(\mu+p)} \sum_{\ell=0}^{N-1} \Bigl( \frac{p}{\mu+p} \Bigl)^\ell
	\biggl[ \mu t + F_\Gamma(t; \ell+1,\mu+p) -
	\frac{\mu}{\mu+p} \sum_{m=0}^\ell F_\Gamma(t; m+1,\mu+p) \biggl].
\end{equation*}

\noindent We can now expand the distribution functions in terms of a
finite sum. Performing some algebraic simplifications
yields the closed form representation

\begin{eqnarray}
\label{Q-t}
\nonumber \fl q(t) &=& 1 - \frac{1}{t(\mu+p)} \sum_{\ell=0}^{N-1} \Bigl(
\frac{p}{\mu+p} \Bigl)^\ell \biggl[ \frac{p - \ell \mu}{\mu+p} + \mu
  t + \\
\fl
&& \qquad \qquad e^{-(\mu+p)t} \sum_{m=0}^\ell \biggl( \frac{\mu}{\mu+p}
  \sum_{i=0}^m \frac{((\mu+p)t)^i}{i!}  - \frac{((\mu+p)t)^m}{m!}
  \biggl) \biggl].
\end{eqnarray}

\noindent
Eq.\,\ref{Q-t} may be further simplified by extracting the first two
terms in the brackets from the sum and using the identity

\begin{equation}
\sum_{\ell=1}^n \ell a^\ell = \frac{a(1 - a^{n+1}) - (n+1)(1-a)a^{n+1}}{(1-a)^2},
\end{equation}

\begin{figure}[t]
\begin{center}
\includegraphics[width=3.5in]{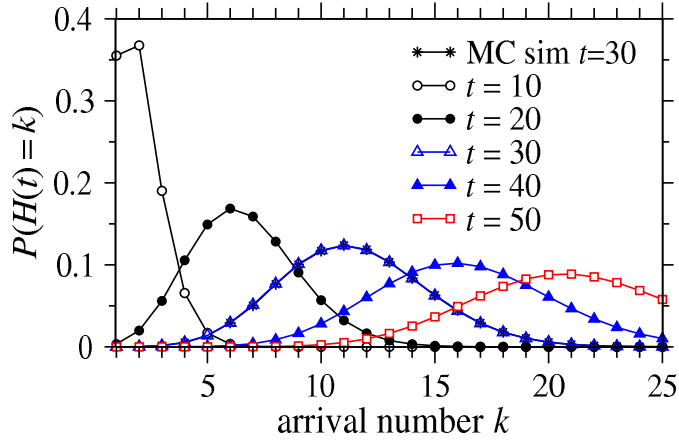}
\caption{Plots of the distribution $P(H(t)=k)$ of the number of hits
$H(t)$ that have occured up to time $t$.  Parameters used were
$p\equiv 1$, $N=10$, $\mu=0.2$, and $\alpha = 3$.  The distribution is
plotted for times $t=5,10,20,30,40,50$.  Monte Carlo simulation
(300000 runs) was used to verify the results for $t=30$.}
\label{H}
\end{center}
\end{figure}

\noindent obtained by differentiating the well-known identity
$\sum_{\ell=0}^n a^\ell = (1 - a^{n+1})/(1 - a)$.  We finally end up
with

\begin{eqnarray} \fl
q(t) = \Bigl(\frac{p}{\mu+p}\Bigl)^N 
		  \Bigl( 1 - \frac{N}{t(\mu+p)} \Bigl) 
		  - \frac{1}{t(\mu+p)} \nonumber \\
\label{PARAMETER_2} \fl
\quad \quad \times \, e^{-(\mu+p)t} \, \sum_{\ell=0}^{N-1} 
\Bigl( \frac{p}{\mu+p} \Bigl)^\ell \, \sum_{m=0}^\ell \biggl( 
\frac{\mu}{p + \mu} \sum_{i=0}^m \frac{((\mu+p)t)^i}{i!}  
- \frac{((\mu+p)t)^m}{m!} \biggl),
\end{eqnarray}

\noindent which yields Eq.\,\ref{PARAMETER}.  In Fig. \ref{H}, and in
all subsequent plots, we nondimensionalise all rates in terms of $p$
and times in terms of $p^{-1}$.  The distribution $P(H(t)=k)$ plotted
in Fig. \ref{H} shows that the probability of more arrivals increases
with time after the start of injection increases.  We compared and
verified our results at time $t=30$ with a Monte Carlo simulation of
the bulk dynamics using the Bortz-Kalos-Lebowitz algorithm
\cite{BKL1975}.  We performed $300000$ runs, each with parameters
$t=30$, $N=10$, $p=1$, $\mu=0.2$ and $\alpha = 3$.





From the survival probability $S_{k}(t)$ found from Eq.\,
\ref{SURVIVE}, all moments $\sigma$ of the $k^{\rm th}$ passage times
can be computed,


\begin{equation}
\label{ESK}
\langle T^{\sigma}_k\rangle = -\int_0^{\infty} t^{\sigma} \frac{\textrm{d}S_k}{\textrm{d}t} \dd t. 
\end{equation}

\noindent The mean ($\sigma = 1$)  {\it first} ($k=1$) passage time to 
the absorbing site can be found from 

\begin{eqnarray} \fl
S_1(t) = P(H(t) = 0) = \exp \biggl[ \alpha
\Bigl(\frac{p}{\mu+p}\Bigl)^N \Bigl(\frac{N}{\mu+p} - t \Bigl)
\nonumber \\
\label{SURVIVAL_PROB} \fl
\quad + \frac{\alpha}{\mu+p} \, e^{-(\mu+p)t} \, \sum_{\ell=0}^{N-1} 
\Bigl( \frac{p}{\mu+p} \Bigl)^\ell \, \sum_{m=0}^\ell \biggl( 
\frac{\mu}{\mu+p} \sum_{i=0}^m \frac{((\mu+p)t)^i}{i!}  
- \frac{((\mu+p)t)^m}{m!} \biggl) \biggl]
\end{eqnarray}
and Eq.\,\ref{ESK}. An explicit expression can be found for
a single-site ZRP ($N=1$):

\begin{eqnarray}
\label{FPTGAMMA}
\langle T_1 \rangle =  {e^{x}x^{-x}\over \mu+p}\gamma(x,x), \quad N=1,
\end{eqnarray}

\noindent where 

\begin{equation}
x\equiv {\alpha p\over (\mu+p)^{2}}. 
\end{equation}
Upon approximating the lower incomplete gamma function $\gamma(x,x)
\equiv \Gamma(x) - \Gamma(x,x)$ in the $x \to 0$ limit, we find

\begin{eqnarray}
\langle T_1 \rangle = \frac{1}{\mu + p}
\left[{1\over x} + 1 + \OO(x)\right], \quad N=1.
\end{eqnarray}

\noindent In the $x \rightarrow \infty$ limit, we apply the  method of steepest
descents \cite{BENDER} to the integral definition of
the $\Gamma$-function to find

\begin{eqnarray}
\langle T_1 \rangle = {1\over \mu + p}\sqrt{\frac{\pi}{2x}}
\left[1 + {12\over x} + \OO(x^{-2})\right], \quad N=1.
\end{eqnarray}

\noindent
For $N > 1$, the integral in Eq.\,\ref{ESK} does not have a simple
representation in non-integral form, nor can the mean $k^{\rm th}$ passage
times be calculated explicitly in the $N=1$ case for $k > 1$.  However
we can find some asymptotic results for large $k$ values in the ``fast
dynamics" limit.  If we denote by $\tau$ the characteristic time $\tau
= N/(\mu+p)$ for a particle just injected to reach the final site, and
consider times $t$ such that $t \gg \tau$ Eq.\,\ref{PARAMETER} may be
written as

\begin{equation}
\lambda(t) = \alphaeff (t - \tau) + \OO(e^{-t(\mu+p)}),
\label{LAMBDA}
\end{equation}
where $\alphaeff = \alpha p^N (\mu+p)^{-N}$ is an effective injection rate 
from the perspective of the final site that takes into account decay.
Because $t \gg \tau$ holds for all but a negligible part of the range of
the integral in Eq.\ref{ESK}, we have

\begin{eqnarray}
\label{T_K}
\langle T_k \rangle = \frac{k}{\alphaeff} + \tau + \OO(\tau^2).
\end{eqnarray}
\noindent
The assumption on $t$ translates onto a condition on $k$, so that
Eq.\,\ref{T_K} remains valid as long as $\langle T_k \rangle \gg
\tau$, or $k \gg \alphaeff \tau $.  

\begin{figure}[t]
\begin{center}
\includegraphics[width=5in]{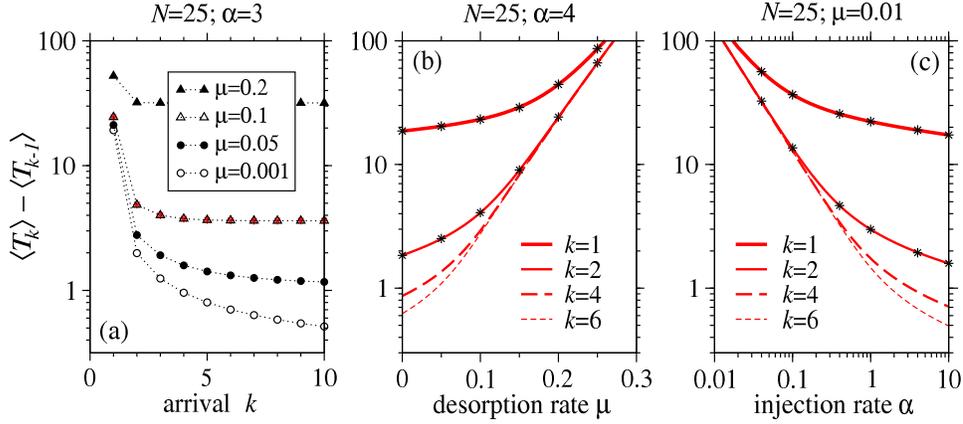}
\caption{Interarrival times $\langle T_{k}\rangle - \langle
T_{k-1}\rangle$ for the ZRP obeying bulk dynamics (with $p\equiv 1$).
(a) Interarrival times as a function of arrival $k$ various values of
the desorption rate $\mu$ for fixed $\alpha = 3$. (b) Arrival times of
the first, second, fourth, and sixth particles as a function of $\mu$
for fixed $\alpha = 4$. (c) First, second, fourth, and sixth arrival
times as a function of the injection rate $\alpha$ for $\mu=0.01$.
All plots were evaluated using $N=25$.  Results were verified
using Monte-Carlo simulations (asterisks).}
\label{DELTAT}
\end{center}
\end{figure}

In Fig. \ref{DELTAT}, we plot the interarrival times $\langle T_{k}
\rangle - \langle T_{k-1} \rangle$ as a function of $k$, $\mu$, and
$\alpha$.  Fig. \ref{DELTAT}(a) shows that larger desorption rates permit
the system to reach steady-state faster, where interarrival times
approach the limit $1/\alphaeff$, for smaller values of $k$. In
Fig. \ref{DELTAT}(b), we see that the mean interarrival times,
including the mean first passage time, increase exponentially for
large $\mu$.  This result is different from that of the problem where
only a single particle is injected, in which the {\it conditional}
mean first passage time of that single particle {\it decreases} has
desorption $\mu$ increases.  Because the single particle problem needs
to be conditioned on arrival, only very fast trajectories can survive
the desorption process, leading to mean first passage times that
decrease rapidly with increasing $\mu$.  Finally, Fig. \ref{DELTAT}(c)
plots the mean interarrival times as a function of injection rate
$\alpha$.

In the next section we rederive the same results above by solving the
corresponding Master equation using generating functions.  Using this
approach, we can not only recover the mean passage times to site
$N+1$, but the full particle occupation distribution function $P(n_1,
\dots, n_{N+1},t)$.

\subsection{Solving the bulk dynamics via generating functions}
\label{BULK_GF}

\noindent
In this section, for mathematical completeness, we rederive the
survival probability using generating function methods applied to the
Master equation for describing the probability $P(\{ n_\ell \},t)$ of
having $\{n_\ell\}$ particles on each of the $1 \leq \ell \leq N+1$
sites:

\begin{eqnarray}
\label{MASTEREQ}
\fl \dot{P}(\{n_{\ell}\}, t) &=& -\alpha \,
[P(\{n_{\ell}\}, t) - P(n_1 -1, \dots, n_{N+1},t ) 
(1 - \delta_{n_1,0})] \\
\fl
\nonumber
&& - (\mu+p)
\sum_{j=1}^{N} n_j \,P(\{n_{\ell}\},t)  + \mu \sum_{j=1}^{N} 
(n_j + 1) P(n_1, \cdots, n_{j} +1, \cdots,t) \\
\fl
\nonumber
&& + p \sum_{j=1}^{N} 
(n_j + 1) P(n_1, \cdots, n_{j} +1, n_{j+1} -1 \cdots,t)(1-\delta_{n_{j+1},0}).
\end{eqnarray}

\noindent
The survival probability $S_1(t)$ is defined by the probability of
having no particles in the absorbing site

\begin{eqnarray}
S_1(t) = \sum_{n_1, n_2 \cdots n_{N}} P(n_1, \cdots, n_{N+1} = 0, t).
\end{eqnarray}

\noindent After setting $n_{N+1} = 0$ in Eq.\,\ref{MASTEREQ}, we
consider the dynamics of the constrained generating function defined
as

\begin{eqnarray}
\fl
G_0(z_1, \cdots, z_{N},t) = \sum_{n_1,\cdots n_{N}=0} ^{\infty}
z_1 ^{n_1} \cdots z_{N}^{n_{N}}P(n_1,\cdots n_N, n_{N+1}=0,t).
\end{eqnarray}

\noindent Upon multiplying  Eq.\,\ref{MASTEREQ} 
by $z_{1}^{n_{1}}\cdots z_{N}^{n_{N}}$ and 
summing over all possible values of $n_j$, $1
\leq j \leq N$, we find a first order partial differential equation for 
$G_{0}(z_{1},\ldots,z_{N},t)$ which we can solve using 
the method of characteristics \cite{WHITMAN}.
We find that $G_{0}$ obeys 

\begin{eqnarray}
\label{longg}
\frac{\dd G_0}{\dd t} &=
& \alpha  (z_1 - 1) G_0
\end{eqnarray}

\noindent along the trajectories defined by

\begin{equation}
\label{TRAJECTORIES}
\begin{array}{l l l}
\displaystyle{\frac{\dd z_1(t)}{\dd t}} &=&
(\mu + p) z_1(t) - \mu - p z_2(t), \\
\nonumber
\\
\displaystyle{\frac{\dd z_j(t)}{\dd t}} &=&
(\mu + p) z_j(t) - \mu - p z_{j+1}(t), \\
\nonumber \\
\nonumber
\displaystyle{\frac{\dd z_{N}(t)}{\dd t}} &=&
(\mu + p) z_{N}(t) - \mu. \\
\nonumber
\end{array}
\end{equation}

\noindent The initial condition $P(n_1, \cdots, n_N, n_{N+1}=0, t=0)=
\delta_{n_1,0} \cdots \delta_{n_N,0} $ gives $G_0(z_0, \cdots z_{N},
t=0) = 1$.  Equations \,\ref{TRAJECTORIES} can be written in the 
form

\begin{equation}
{\dd {\bf Z}(t) \over \dd t} = {\bf M}{\bf Z}(t) - \mu {\bf I},
\label{tosolve}
\end{equation}
where ${\bf Z}(t) = (z_{1}(t), \ldots,z_{N}(t))^{T}$ is the vector of
trajectories, ${\bf I}$ is the $N\times N$ identity matrix,
and ${\bf M}$ is a tridiagonal matrix with elements $m_{j,j} = \mu +
p$, $m_{j,j+1} = -p$, and $m_{i,j}=0$ otherwise.
Upon defining the Laplace transform $\tilde{{\bf
Z}}(s) = \int_{0}^{\infty}{\bf Z}(t) e^{-st}\dd t$ and the initial
values ${\bf Z}(t=0) = (z_1(t=0), \cdots, z_N(t=0))^{T}$, 
Eqs.\,\ref{tosolve} can be written in the form

\begin{equation}
s \tilde{{\bf Z}}(s) = {\bf M} \tilde{{\bf Z}}(s) -\frac {\mu}{s}{\bf I} + {\bf Z}(t=0),
\end{equation}
and solved explicitly by first inverting $s{\bf I} - {\bf M}$ and then
calculating the inverse Laplace transform of $\tilde Z(s)$.
After performing the algebra, the solution to Eqs.\,\ref{tosolve}
can be expressed as

\begin{equation}
\label{forward}
(z_j - R_{N-j}) = \sum_{k=0}^{N-j} 
(z_{j+k}(t=0) - R_{N-j-k})
\frac{(-p t)^k}{k!} e^{(\mu +p)t},
\end{equation}

\noindent where 

\begin{equation}
R_{k} \equiv 1 - \left(\frac{p}{\mu+p}\right)^{k+1}\!\!\!.
\nonumber
\end{equation}

\noindent Upon using $z_{1}(t)$ from Eq. \ref{forward}
in Eq. \ref{longg}, we find $G_{0}$ as a function of 
the $z_j$ values implicitly expressed through the starting 
positions $z_{j}(t=0)$ of the trajectorie:

\begin{eqnarray}
\fl
G_0(z_1, \cdots, z_N, t) = \exp
\left[ - \alpha t
\left(\frac{p }{\mu + p}
\right)^{N}- \right. && \left.\alpha \sum_{k=0}^{N-1} 
{(z_{k+1}(t=0) - R_{N-k-1})p^{k}\over (\mu+p)^{k+1} k!}
\right.
\\ 
\nonumber
&&
\left.
\frac{}{}^{}
\times 
\gamma[k+1, - (\mu + p) t] \,
\right].
\label{G0}
\end{eqnarray}
We are thus left with explicitly determining
$z_j(t=0)$ as a function of the independent
variables $z_j$. We do this by
inverting Eq.\,\ref{forward}

\begin{eqnarray}
\label{backward}
(z_j(t=0) - R_{N-j}) = \sum_{k=0}^{N-j} (z_{j+k} - R_{N-j-k}) \frac{(p
  t)^k}{k!} e^{-(\mu +p)t}.
\end{eqnarray}

\noindent Using this result in Eq. \ref{G0} we find

\begin{eqnarray}
\fl
G_0(z_1, \cdots, z_N, t) = \exp
\left[ - \alpha t
\left(\frac{p }{p + \mu}
\right)^{N}
\right. - \alpha
e^{-(\mu + p)t}
\sum_{k=0}^{N-1} 
\sum_{j =0}^{N-k-1} 
\frac{ p^{j+k} t^{j} } 
{(\mu+p)^{k+1} k! j!} \\ 
\nonumber
\hspace{2cm} \times (z_{j+k+1}- R_{N - j - k -1})
\label{fullgenerate}
\left.
\gamma[k+1, -(\mu+p) t] \,
\frac {}{}
\right].
\end{eqnarray}

\noindent
Finally, since the survival probability
is obtained by imposing $z_{\ell} = 1$ for all $\ell$,
we obtain

\begin{eqnarray}
\fl
\nonumber
S_1(t) = \exp
\left[ - \alpha t 
\left(\frac{p }{p + \mu}
\right)^{N} \right. \\ 
\hspace{0cm}
\left. - \frac{ \alpha p^N e^{-(\mu+p)t}}
{(\mu + p)^{N+1}}
\sum_{j=0}^{N-1} 
\sum_{\ell =0}^{N-1-j} 
\frac{t^{\ell}
(\mu + p)^{\ell}}
{j! \ell !} 
\gamma[j+1, - (\mu + p) t] \,
\right],
\end{eqnarray}

\noindent
which is equivalent to Eq.\,\ref{SURVIVAL_PROB}.  We can now
successively determine the dynamics of the probability distribution
function conditioned on the absorbing site containing a finite number
of particles $n_{N+1} \geq 1$. For $n_{N+1} = 1$, the corresponding
generating function $G_1(z_1, \cdots, z_N,t)$ can be obtained from the
Master equation for $P(n_1, \cdots, n_{N+1} =1,t)$:

\begin{eqnarray}
\fl
G_1(z_1, \cdots, z_N,t) = \sum_{n_1, \cdots n_N=0}^{\infty}
z_1^{n_1} \cdots z_N^{n_N} P(n_1, \cdots n_N, n_{N+1} =1,t).
\end{eqnarray}

\noindent
Upon summing Eq.\,\ref{MASTEREQ} over all values of $n_j$
we find that the dynamics of $G_1$ is given by

\begin{eqnarray}
\label{ONELAST}
\frac{\partial G_1}{\partial t} = \alpha G_1(z_1 -1) + \frac{\partial G_0}{\partial z_1},
\end{eqnarray}

\noindent
where $G_0$ is the generating function associated with the adsorbing
site having no particles, $n_{N+1} =0$, and where the evolution of the
trajectories $z_{j}(t)$ are unchanged from those described by
Eqs.\,\ref{TRAJECTORIES}.  The solution to Eq.\,\ref{ONELAST} can be
expressed in the form

\begin{eqnarray}
G_1(z_1, \cdots, z_N,t) =  \lambda(t) \,G_0(z_1, \cdots, z_N,t),
\end{eqnarray}

\noindent where $\lambda(t)$ obeys 

\begin{eqnarray}
\frac{\dd \lambda(t)}{\dd t} = \frac p {G_0} \frac{\partial G_0}{\partial z_1}.
\end{eqnarray}

\noindent The solution for $\lambda(t)$ turns out to be precisely that
given in Eq.\,\ref{PARAMETER}.  Similarly, it can be found that the
generating function with the constraint $n_{N+1} = j$ is given by

\begin{eqnarray}
\nonumber
\label{GENERALJ}
\fl
G_{j}(z_1, \cdots, z_N,t) &=& 
\sum_{n_1, \cdots n_N=0}^{\infty}
z_1^{n_1} \cdots z_N^{n_N} P(n_1, \cdots n_{N+1} =j,t) \\
&=&
\frac{\lambda(t)^{j}}{j!} G_0(z_1,\cdots, z_N,t),
\end{eqnarray}

\noindent
The survival probability $S_{k}(t)$ is given by $S_k(t) =
\sum_{j=0}^{k-1} G_{j}(1, \cdots, 1,t)$ and moments of the $k^{\rm th}$
arrival times, found previously, can also be obtained using
Eq.\,\ref{ESK}.  In addition, an advantage of the generating function 
approach is that the particle occupations can 
also be determined. For example, the mean occupation at site $\ell$, 
conditioned on exactly $j$ particles having entered site $N+1$ is
given by 

\begin{eqnarray}
\fl
\nonumber
\langle n_{\ell}(t\vert n_{N+1} = j)\rangle &=
\sum_{n_1, \cdots, n_{N}=0} ^{\infty} n_\ell P(n_1, \cdots, n_{N+1}=j,t) 
\nonumber \\
& = \frac{\lambda(t)^j}{j!} 
\frac{\partial G_0(1,\cdots, z_{\ell}, \cdots, 1,t)}{\partial z_\ell} \bigg|_{z_\ell =1} 
\nonumber \\ 
& = - \frac{\lambda(t)^j}{j!} e^{-\lambda(t)}
\alpha p^{\ell-1} e^{-(\mu+p) t}
\sum_{k=0}^{\ell -1} 
\frac{\gamma[k+1, -(\mu+p)t] \, t^{\ell -k-1}}{(\mu + p)^{k+1} k! (\ell -k-1)!}.
\label{ABOVE}
\end{eqnarray}


\noindent
Upon summing Eq.\,\ref{ABOVE} over all $j$, we find that the
unconditioned mean occupation $\langle n_{\ell} (t) \rangle$
is given by

\begin{eqnarray}
\fl
\langle n_{\ell}(t)\rangle &=& 
\hspace{-3mm}\sum_{n_1, \cdots, n_{N+1} =0} ^{\infty} n_\ell P(n_1, \cdots, n_{N+1},t) \\
\nonumber 
\fl
&=& - \alpha p^{\ell-1} e^{-(\mu+p) t}
\sum_{k=0}^{\ell -1} \frac{\gamma[k+1, -(\mu+p)t] 
\,t^{\ell -k-1}} {(\mu + p)^{k+1} k! (\ell -k-1)!}.
\label{MEAN_OCCUPANCY}
\end{eqnarray}

\noindent 
Two limits are of interest: the mean occupation conditioned on no
particles hitting site $N+1$, which is given by

\begin{eqnarray}
\langle n_{\ell}(t\vert n_{N+1}=0) \rangle &=& 
e^{-\lambda(t)} \, \langle n_{\ell} (t) \rangle, 
\end{eqnarray}

\noindent
and the average occupation of site $N+1$, regardless of the occupation
state of all other sites, which is simply $\langle n_{N+1} (t)\rangle
= \lambda(t)$. Thus, in the long time limit, the occupation of the
final site $N+1$ will scale as

\begin{eqnarray}
\langle n_{N+1} (t)\rangle \sim \alpha t \left(\frac{p}{\mu+p}\right)^{N}, 
\end{eqnarray}

\noindent indicating that at site $N+1$ particles accumulate linearly
at a rate that is proportional to the injection rate $\alpha$
attenuated by the evaporation probability for each of the $N$
intervening sites.

\begin{figure}[t]
\begin{center}
\includegraphics[width=4.2in]{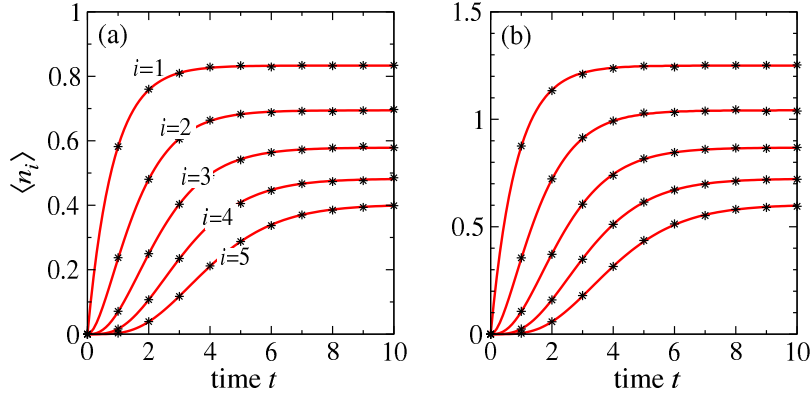}
\end{center}
\caption{Time dependence of the mean site occupancies.  Both panels
display exact values (solid lines) from Eq.\, \ref{MEAN_OCCUPANCY} and
simulation (asterisks) for parameter values $p\equiv 1$, $N=5$, and
$\mu=0.2$.  Each curve and approximating points correspond to mean
occupations at different sites, with earlier site having higher
occupations.}
\label{DENSITY}
\end{figure}

In Figure \ref{DENSITY}, we have plotted the mean occupations derived
from Eq.\,\ref{MEAN_OCCUPANCY} for $N=5$ and $\mu=0.2$.  All mean
occupancies are seen to reach steady state by $t \approx 10$ and, for
all times, mean occupancy is a monotonically decreasing function of
site index due to decay.  In Fig. \ref{DENSITY}(a), $\alpha = 1 < p +
\mu$, implying that particles are cleared out faster than they are
injected, resulting in $\langle n_{1}\rangle$ approaching a value
less than one.  In (b), the $\alpha = 1.5 > p + \mu$, and $\langle
n_{1}\rangle$ (and $\langle n_{2}\rangle$ asymptotes to values greater
than one. Our results are verified with Monte Carlo simulations.

Finally, the full distribution for $P(n_1,\cdots, n_{N+1}=j,t)$ can be
found by using the Cauchy integral \cite{CAUCHY} over
Eq.\,\ref{GENERALJ}

\begin{eqnarray}
\fl P(n_1, \ldots, n_{N+1}=j,t) 
= \frac 1 {2 \pi i} \oint_C
\frac{G_j(z_1, \cdots, z_N, t)}{z_1^{n_1 +1}, \cdots z_N^{n_N+1}}
\dd z_1 \cdots \dd z_N,
\end{eqnarray}

\noindent
where the integral is closed along a path encircling the
origin. Evaluating the residues, the above integral can be expressed
as

\begin{eqnarray}
\fl P(n_1, \ldots, n_{N+1}=j,t) = 
\frac{\lambda(t)^j}{j!} \prod_{\ell=1}^{N} 
\left(\frac{\partial}{\partial z_\ell}\right)^{n_{\ell}} G_0(z_1, \cdots, z_N,t)\big|_{z_{\ell} = 0},
\end{eqnarray}

\noindent
which can be calculated explicitly to yield

\begin{eqnarray}
\fl
\nonumber
P(n_1, \ldots, n_{N+1} =j,t) = \frac{\lambda(t)^j G_0(0,\cdots, 0, t)}
{j!} \\
\hspace{0.3cm}\times \prod_{\ell=1}^{N} 
\left[ -\alpha e^{-(\mu+p) t}
  \sum_{k=0}^{\ell-1} \frac{p^{\ell-1} t^{\ell-1-k}}{(p + \mu)^{k+1}
    k!} \gamma[k+1, -(\mu+p) t]\right]^{n_{\ell}}\!\!\!\!.
\label{PTOTAL}
\end{eqnarray}

\section{Zero-range Model with Surface Dynamics}


Surface dynamics differs from bulk dynamics in that only the top
particle at a given site is able to hop or decay, while the ones below
remain inert.  The dependencies thus introduced between the particles
renders the $k$th hitting time, and even the survival probability
$S_k(t)$ of the final site, too difficult to derive in the general $N$
case.  In particular, we cannot use the strategy employed for bulk
dynamics because it relied essentially on particles injected before
time $t$ having {\it independent} probabilities of reaching site $N+1$
by time $t$.  In the case of surface dynamics, only the top particle
in a pile attempts to move to a neighboring pile. The difficulty
arises in keeping track of which sites are empty and which ones contain
at least one particle.

Beginning with the approach of the previous sub-section, we first
consider the Master equation for the distribution of the site
occupancies obeying surface dynamics

\begin{eqnarray}
\label{SURFACEDYN}
\fl \dot{P}(\{n_\ell\},t) &=& 
- \alpha [ P(\{ n_{\ell} \},t) -P(n_1 - 1, \ldots, n_{N+1},t) 
(1 - \delta_{n_1,0})] \\
\nonumber
\fl 
&& -
\mu \sum_{j=1}^N [ (1 - \delta_{n_j,0}) 
P(\{n_\ell\},t) - P(n_1, \ldots, n_{j}+1, \cdots, t)]
\\
\nonumber
\fl
&& - p \sum_{j=1}^N (1 - \delta_{n_j,0}) 
P(\{n_\ell\},t)  \\ \nonumber
&& - p\sum_{j=1}^{N} - (1 - \delta_{n_{j+1,0}})P(n_1, \ldots, n_{j} + 1, n_{j+1}-1, \ldots, t).
\end{eqnarray}

\noindent We now introduce the marginal probability

\begin{eqnarray}
P_{i}(n_i,t) = \sum_{\{n_{j\neq i}\}=0}^{\infty} 
P(n_1, \ldots, n_{N+1},t),
\label{PMARG}
\end{eqnarray}
where the sum is taken over all $n_{j}$ for all sites $1\leq j \leq
N+1$, except site $j=i$.  Eq.\,\ref{PMARG} represents the probability
that site $i$ has $n_{i}$ particles regardless of the occupation of
all other sites.  Similarly, the joint probability for sites $i-1$ and
$i$ is defined as

\begin{eqnarray}
P_{i-1, i}(n_{i-1}, n_{i},t) = \sum_{n_{j\neq i-1,i} =0}^{\infty} 
P(n_1, \ldots, n_{N+1},t).
\end{eqnarray}

\noindent
Upon summing Eq.\,\ref{SURFACEDYN} over all values of $n_{j\neq i}$,
we find the time evolution for the marginal probability $P_i(n_i,t)$
as a function of the two-site probabilities $P_{i-1, i}(n_{i-1},
n_{i},t)$:

\begin{eqnarray}
\fl
\label{SECOND3}
\dot P_i(n_i, t) &=& p \sum_{n_{i-1} =1}^{\infty} (1 - \delta_{n_i,0})
P(n_{i-1}, n_i -1,t) - p \sum_{n_{i-1}=1}^{\infty}
P(n_{i-1}, n_i,t) + \\ \nonumber
\fl
&& + (\mu+p) P_i(n_i +1,t) - (\mu+p) P(n_i,t) (1 - \delta_{n_i,0}), \quad\,\, 1<i\leq N+1 \\
\end{eqnarray}
Continuing in this way, the equations for the marginal occupation
probabilities form a hierarchy which is completed by the equation for
the injection site $i=1$:

\begin{eqnarray}
\label{FIRST}
\dot P_{1}(n_1,t) &=& - \alpha [P_{1}(n_1,t) - P_{1}(n_1 -1,t) (1 - \delta_{n_1,0})] \\
\nonumber
\fl 
&&- (\mu + p) [P_{1}(n_1,t) (1 - \delta_{n_1,0}) - P_{1}(n_1 +1,t)].\\
\fl
\nonumber
\end{eqnarray}

\noindent 
Note that the dynamics for site $i=1$ is completely decoupled from that
of the other sites so that the marginal occupation distribution of the
first site can be solved directly. We now consider two cases where 
analytic results can be found.

\subsection{Single-site ZRP densities and mean first passage times}

Since Eq.\,\ref{FIRST} is decoupled from the hierarchy, 
we can solve it by taking its Laplace transform and using the initial 
condition $P(n_1,0) = \delta_{n_1,0}$ to find

\begin{equation}
\begin{array}{l}
\displaystyle \Pt_{1}(n_{1}=1,s) = {s+\alpha \over
  \mu+p}\Pt_{1}(n_{1}=0,s) - {1\over \mu + p} \\[13pt] \displaystyle
\Pt_{1}(n_{1}+1,s) = \left(1+{s+\alpha \over \mu +
  p}\right)\Pt_{1}(n_{1},s) - {\alpha \over \mu+p}\Pt_{1}(n_{1}-1,s).
\end{array} 
\end{equation}

\noindent The solution can be expressed in the form 

\begin{equation}
\label{SOLVESURFACE}
\Pt_{1}(n_{1},s) = \left[\frac{1-z_1(s)} {s} \right] z_1(s)^{n_{1}}, 
\label{P_1}
\end{equation}

\noindent
where

\begin{equation}
\fl
z_1(s) = \frac{1}{2(\mu+p)} \left( (s+\alpha+\mu+p) 
- \sqrt{(s+\alpha+\mu+p)^2 - 4 \alpha (\mu+p)} \right).
\end{equation}
Upon inverting, we find 

\begin{eqnarray} 
z_1(t) 
&=& \frac{e^{-(\alpha+\mu+p)t}\sqrt{\alpha}}{t\sqrt{\mu+p}} I_1( 2 \sqrt{\alpha
  (\mu+p)} t),
\label{ZSOL}
\end{eqnarray}

\noindent
where $I_1(t)$ is the first-order modified Bessel Function of the
first kind.
From Eq.\,\ref{SOLVESURFACE} and using the fact that the inverse
Laplace transform of a product is a convolution in time,
we can iteratively construct $P_1(n_1,t)$ starting from $P_1(0,t)$

\begin{eqnarray}
\label{CONVOLUTE1}
P_1(n_1=0,t) = 1 - \int_0^t z_1(t') \dd t',
\end{eqnarray}

\noindent where $z_1(t)$ is given by Eq.\,\ref{ZSOL} and

\begin{eqnarray}
\label{CONVOLUTE2}
P_1(n_1,t) = \int_0^{t} P_1(n_1-1,t') z_1(t') \dd t'.
\end{eqnarray}

\noindent
The integrals in Eq.\,\ref{CONVOLUTE1} and \ref{CONVOLUTE2} do not
have simple closed forms. However, the functions $P_1(n_1,t)$ can also
be obtained from differentiation using the relation

\begin{eqnarray}
\fl
P_1 (n_1+1, t)&= \left(1-\delta_{n_{1},0} + \frac{\alpha}{\mu+p} \right) 
P_1(n_1,t)  \\ \nonumber 
& \hspace{1cm} +\frac{\dot  P_1(n_1,t) - \alpha P_1(n_1-1,t) (1 - \delta_{n_1,0})}
{\mu+p}.
\end{eqnarray}

\noindent
For instance, we may recursively write

\begin{eqnarray}
P_1(n_1=1,t) = \frac{\alpha}{ \mu+p} 
\left(1 - \int_{0}^t z_1(t') \dd t' \right) - \frac{z_1(t)}{\mu+p}.
\end{eqnarray}
In the case of $N=1$ we can also solve for the {\it first} ($k=1$)
passage times by observing that the equation for the two-site
distribution function, conditioned on $n_{2}=0$, is also decoupled from
the hierarchy:

\begin{eqnarray}
\fl
\nonumber
\dot P(n_1,0,t) &=& \alpha [P(n_1-1,0,t) (1- \delta_{n_1,0}) - P(n_1,0,t)] + 
\mu P(n_1+1,0,t)  \\
\fl
&& - (\mu +p) P(n_1,0,t)) (1- \delta_{n_1,0}),
\nonumber
\end{eqnarray}
where we have dropped the subscripts on the two-site distribution
function so that $P(n_{1},n_{2},t) \equiv P_{1,2}(n_{1},n_{2},t)$.
Using Laplace transforms, we find

\begin{eqnarray}
\tilde P(n_1,0,s) = \frac{y_1(s)^{n_1}}{s + \alpha - \mu y_1(s)},
\label{P12}
\end{eqnarray}

\noindent
where

\begin{eqnarray}
y_1(s) = \frac{\alpha + \mu + p + s - \sqrt{(\alpha + \mu + p +s)^2 -
    4 \alpha \mu}} {2 \mu}.
\end{eqnarray}
From this solution of $\tilde{P}(n_1,n_2=0,s)$, we can obtain the
Laplace transformed probability that site $i=2$ has not been hit by a
particle $\tilde{S}_1(s) = \sum_{n_1=0}^\infty
\tilde{P}(n_1,n_2=0,s)$. Thus, in the $N=1$ case, the mean first
passage time is

\begin{eqnarray}
\fl
\label{FPTsurface}
\langle T_1 \rangle &=& \sum_{n_1=0}^{\infty} \tilde P (n_1,0,s=0) =
\frac{\alpha + \mu + p + \sqrt{(\alpha + \mu + p)^2 - 4 \alpha \mu}}
{2 \alpha p}. \\
\fl \nonumber
\end{eqnarray}
Note that in the case of $\mu=0$, this result simplifies to $\langle
T_1 \rangle = \alpha^{-1} + p^{-1}$. In surface dynamics without
desorption, the first passage time is determined by the dynamics of
the lead particle. Therefore, the mean first arrival time is simply
the total time it takes for the leading particle to reach site $N+1$
and is given by $\langle T_1 \rangle = \alpha^{-1} + N p^{-1}$.

\subsection{Steady-state limit}

We have not been able to find closed-form solutions of the surface
dynamics ZRP for general $N$ and nonzero desorption rate $\mu > 0$.
However,  Eq.\, \ref{FIRST} can be solved in the steady state limit by using
the {\it ansatz} $P_1(n_1) = (1 - z_1) \, z_1 ^{n_1}$.  The equation
supports a solution when $z_1 = \alpha (\mu+p)^{-1}$, implying

\begin{eqnarray}
\label{P1}
P_1(n_1) = \left( 1 - \frac{\alpha}{ \mu + p} \right) 
\left(\frac {\alpha}{\mu + p}\right)^{n_1}.
\end{eqnarray}

\noindent 
The above expression is correct only for $\alpha < \mu +p$, so that
$0\leq P_1(n_1) \leq 1$.  Physically this condition is simply a
statement of that if injection is too fast, the occupations continues
to build without bound.  Steady-state occupations arise only if the
injection rate $\alpha$ is small enough such that hopping $p$ and
evaporation $\mu$ can keep up.

In order to solve Eq.\,\ref{SECOND3}, we need to a closure relation
for the two-site probability distribution $P_{i-1,i}(n_{i-1}, n_i)$.
As shown in \cite{Zia}, the two-site probability distribution can be
factorised in the steady-state limit and expressed as
$P_{i-1,i}(n_{i-1}, n_i) = P_{i-1}(n_{i-1}) \, P_i(n_i)$.  If we
impose that each $P_j(n_j)$ has a power law dependence in $n_j$,
similar to what done for $n_1$, it is easy to verify that the
steady-state marginal probabilities are solved by

\begin{eqnarray}
\label{pk}
P_j(n_j) = \left(1 - \frac{\alpha p^{j-1} }{(\mu + p)^j} \right) 
\left(\frac {\alpha p^{j-1} }
{(\mu + p)^{j}}\right)^{n_j}, \quad \alpha <
\mu+p.
\end{eqnarray}

\noindent The resulting steady-state mean
occupation at each site are

\begin{eqnarray}
\langle n_j \rangle = \frac {\alpha p^{j-1}}{ (\mu + p)^j - \alpha p^{-1}}, \quad \alpha <
\mu+p.
\end{eqnarray}
From our steady-state results for $N>1$, we can find an approximation
to the passage times by a mean-field argument in which the probability
of site $N+1$ surviving up to $k-1$ particles hitting it obeys
$\dot{S_k}(t) = -J(t)S_k(t)$. The particle current $J(t) = p
\sum_{n_{N}=1}^{\infty}P(n_N | T_k > t)$ is conditioned on fewer than
$k$ particles having arrived at site $N+1$ by time $t$.  Since neither
$P(n_N| T_k > t)$, nor the unconditional distribution $P(n_N,t)$ are
available, we must approximate $J(t)$ with its steady-state,
``mean-field'' (single site marginal distribution) value through
Eq.\,\ref{pk}:

\begin{equation}
\begin{array}{rl}
J & \displaystyle \approx p \sum_{n_{N}=1}^{\infty}P(n_N,t\rightarrow \infty) \\
\: &  \displaystyle  = p \sum_{n_{N}=1}^{\infty}\left(1-{\alpha p^{N-1}\over (\mu+p)^{N}}\right)
\left({\alpha p^{N-1}\over (\mu+p)^{N}}\right)^{n_{N}} \\
\: &  \displaystyle  = {\alpha p^{N}\over (\mu+p)^{N}}.
\end{array}
\end{equation}
In this approximation, $J$ is independent of $k$ and 
all interarrival times are approximately

\begin{equation}
\langle T_{k}\rangle - \langle T_{k-1}\rangle = 
\int_{0}^{\infty}S_{k}(t)\dd t \approx {(\mu+p)^{N}\over \alpha p^{N}}.
\label{approx}
\end{equation}

\begin{figure}[t]
\begin{center}
\includegraphics[width=3.5in]{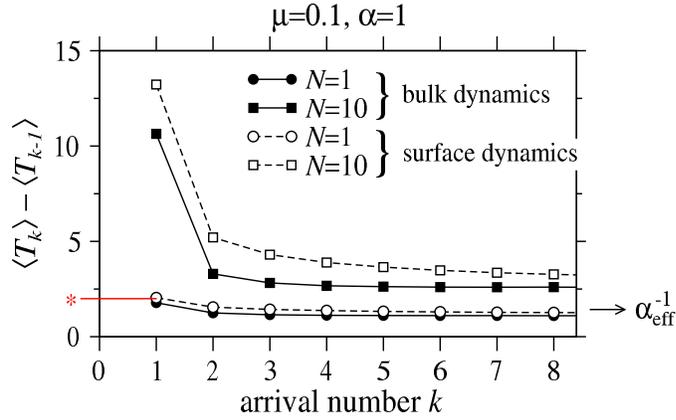}
\caption{Interarrival times $\langle T_{k}\rangle - \langle
T_{k-1}\rangle$ for the surface dynamics ZRP as a function of $k$.
Results from Monte Carlo simulations (open symbols) for both $N=1$ and
$N=10$ are presented. Since fewer particle are mobile in surface
dynamics, the arrival times are longer. The simulations match the
analytic results found for the $N=k=1$ (Eq.\,\ref{FPTsurface},
indicated by the asterisk) and large $k$ (Eqs.\,\ref{approx} and
\ref{T_K}) limits.  For comparison numerical results for the bulk
dynamics case (filled symbols) are also plotted.  All plots were
derived using $\mu=0.1$ and $\alpha = 1$.}
\label{DELTATS}
\end{center}
\end{figure}
As expected, this estimate is precisely that given by $\alpha_{\rm
eff}$ in Eq.\,\ref{LAMBDA} for the bulk dynamics case and is accurate
in the limit of $\alpha \ll (\mu+p)$ where the entry flux is slow
compared to the internal dynamics and the first passage time is
dominated by the contribution given by $\alpha ^{-1}$.  Fast internal
dynamics allows the system to quickly reach steady-state, rendering
the interarrival times equivalent for bulk and surface dynamics.

Upon taking the limit of slow injection rate $\alpha \to 0$ in
Eq.\,\ref{FPTsurface} we find

\begin{eqnarray}
\lim_{\alpha \to 0 }\langle T_1 \rangle = \frac {(\mu+p)}{\alpha p}, 
\end{eqnarray}

\noindent which is identical to the result in Eq.\,\ref{approx} for
$N=k=1$. Figure \ref{DELTATS} plots simulated interarrival times and
compares them with those from bulk dynamics. For large $k$,
interarrival times for both bulk and surface dynamics approach the
same value $\alpha_{\rm eff}^{-1}$ for each $N$. The only exact result
for surface dynamics is that given by Eq.\, \ref{FPTsurface} for
$N=k=1$, and is indicated by the asterisk at $\langle T_{1}\rangle
\approx 2.05125$.

\section{Summary and Conclusions}

In this paper, we have provided detailed and explicit calculations of
first passage times of an $N-$site, one-dimensional Zero-Range
Process.  Both a Poissonian injection process at an injection site,
and spontaneous desorption of all sites were included. We considered
both bulk dynamic and surface dynamic rules as illustrated in
Fig. \ref{Fig1}.

For the ZRP obeying bulk dynamics, we computed the particle passage
times using two methods.  In the first method, we explicitly
enumerated the random walks of each injected particle and evaluated
their probability of reaching the final absorbing site within a time
window. The probability that the absorbing site has not absorbed more
than $k$ particles by a certain time was constructed. The main results
for the survival probabilities are given by Eqs.\,\ref{SURVIVE} and
\ref{PARAMETER}, with explicit expressions for the mean first passage
time given by Eq.\,\ref{FPTGAMMA} and its subsequent asymptotic
limits.

We also derived the complete Master equation for the probability
distribution for a ZRP obeying bulk dynamics, and solved its
corresponding generating function using the method of characteristics.
In addition to the $k^{\rm{th}}$ passage time distribution, this
yielded the mean conditional occupancies of each site given by
Eq.\,\ref{ABOVE}, and the full probability distribution given by
Eq.\,\ref{PTOTAL}.

Finally, for a single site ($N=1$) ZRP obeying surface dynamics, we
found exact results for the site density distribution (Eqs.\,
\ref{CONVOLUTE1} and \ref{CONVOLUTE2}) and the mean {\it first}
passage times (Eq.\,\ref{FPTsurface}).  Note that higher moments of
the first passage time are readily obtained by evaluating higher
derivatives of Eq.\,\ref{P12} at $s = 0$. For general $N$,
only the steady-state particle currents and interarrival times could
be found in closed form (Eq.\,\ref{approx}).

\vspace{5mm} The authors thank T. Antal for fruitful discussions.
This work was supported by the NSF through grants DMS-0349195 and
DMS-0719462 and by the NIH through grant K25AI058672.



\end{document}